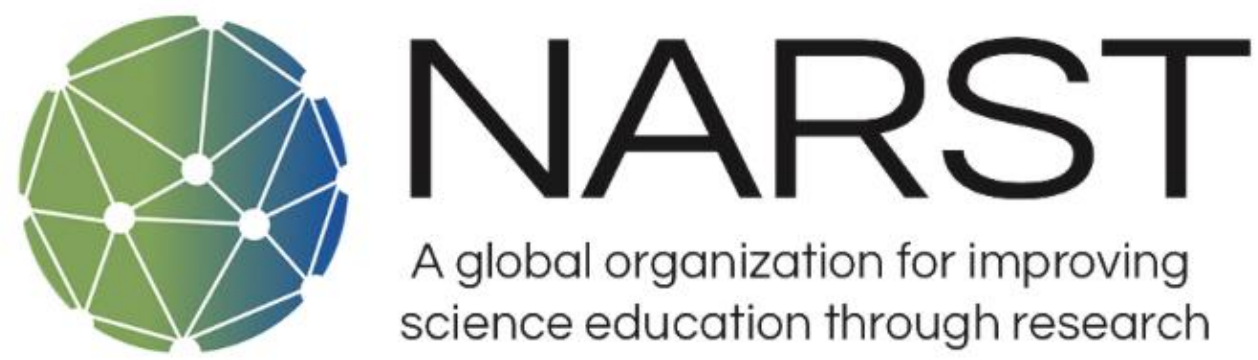

**Concurrent Phenomenological Analysis of STEM Career Aspirations in Underrepresented Youth: Role of Experiences and Identity**

**Author(s):** Amdad Ahmed Awsaf, Florida International University; Nicole Giansanti, Florida International University; Susan Sunbury, Harvard University; Remy Dou, Florida International University; Gerhard Sonnert, Harvard University; Philip Sadler, Harvard University

## Abstract

Despite numerous initiatives aimed at enhancing diversity and achieving equity in the STEM workforce, racially and ethnically minoritized individuals remain underrepresented in STEM disciplines and the STEM workforce. While many factors influence STEM identity (i.e., seeing oneself as a STEM person), it strongly correlates with individuals' future STEM career choices. This phenomenological qualitative study explores the impact of formal or informal STEM-related recognition that influences minority youths' aspirations to pursue a STEM career. Results indicate that though misrecognition negatively impacts one's motivation to study STEM areas, lack of recognition in formal schooling contexts was sometimes mitigated by recognition and support from family. The study suggests providing targeted interventions to facilitate underrepresented youths' achievements in STEM to foster a strong STEM identity and STEM career aspirations.

## Subject/Problem

Chen and colleagues (2024) reported that students' self-perception as scientists positively correlates with their intention to pursue a career in STEM. Carrying out work at an (HSI), Dou and Cian (2021) similarly reported that contextual STEM learning in formal and informal educational settings influences an individual's academic performance, engagement, career choice, and persistence in STEM-related fields. Their findings suggest that negative experiences resulting in a lack of STEM recognition by others could function as a barrier for underrepresented students, impacting their career intentions (Wong et al., 2022). Despite our current understanding, little research has been carried out to map the kinds and scopes of recognition events that contribute to youth STEM identity development, including determining whether or not a set of commonly experienced events can be identified. In this presentation, we will offer preliminary findings from concurrent data analysis pertaining to a phenomenological qualitative study exploring how pre-college formal or informal STEM-related public recognition events (e.g., achievements, failures), as defined by interviewees, may have shaped their identification with STEM fields and career aspirations. This preliminary analysis aims to foster the development of additional hypotheses around the nature of public recognition experiences that could positively shape STEM career aspirations among underrepresented youth (e.g., Cian & Dou, 2024).

Despite numerous initiatives aimed at enhancing social diversity and achieving equity in the STEM workforce (Russo-Tait, 2022), the underrepresentation of women and racialized minorities is still prevalent in the U.S. STEM workforce (Sarraju et al., 2023). As a potential avenue to address these inequities, several studies (e.g., Martin-Hansen, 2018; Griffin, 2018) have pointed to the strong relationship between STEM identity and STEM career aspirations. Contemporary STEM Education research indicates that engagement in STEM-related communities and recreational activities significantly shapes an individual's identification with STEM fields (Dou & Cian, 2021; Verdín et al., 2018) and, relatedly, their career aspirations (Dou et al., 2019). As such, ensuring that youth who engage in STEM learning activities experience settings inclusive of their gender and racial identities could nurture a positive social identification with STEM and career aspirations. However, research indicates that youth from marginalized gender and racial groups tend to experience misrecognition or lack of recognition in STEM contexts—a key suppressor of STEM identity formation (Avraamidou, 2022; Carlone & Johnson, 2007; Rodriguez et al., 2019).

In their framework of "physics identity", Hazari et al. (2010) addressed the concept of discipline-based identity, highlighting three key precursors (i.e., recognition, interest, and performance–competence) that contribute to the development of one's identity. Dou and Cian (2022) extended this work to the context of "STEM identity," finding that "recognition" from others is a critical aspect of STEM identity development (Dou & Cian, 2022). Their research also indicates that students' "performance-competence" in STEM, that is, a person's beliefs about their abilities to successfully perform STEM-related tasks, significantly impacts their STEM identity more than their interest in STEM alone. When taking these three precursors together, students who strongly identify with STEM have been shown to be more than 60% likely to aspire to a STEM career compared with students who do not identify with STEM fields (Dou et al., 2019).

Gee (2000) reported that identity can be developed as a factor of being "recognized as a 'certain kind of person'" (p. 100). Mann (1988) claimed that "recognition" of academic achievement (e.g., honor roll, awards, scholarships, grades) is a positive approach by teachers and parents to acknowledging accomplishments, which might also increase individual performance and sense of accomplishment. Similarly, in general, "recognition" from peers or public entities, such as community organizations, can function as notable contributors to one's enhanced motivation and a sense of accomplishment in STEM contexts (Schmidt et al., 2009). In contrast, a lack of explicit recognition could hinder individuals from aspiring to STEM careers (e.g., Swafford & Anderson, 2020). Li & Singh (2023) reported that due to the general lack of recognition received from others, students in "minority" groups more often exhibit performance-competence in STEM (e.g., MacPhee et al., 2013), which could negatively shape their carer persistence (Rittmayer & Beier, 2008). Avraamidou (2022) asserted that recognition is culture-dependent and often influenced by "cultural and gender stereotypes, policies, racism, sexism, classicism and other forms of discrimination" (p. 85).

Considering the important nature of recognition from others in STEM identity development, we set out to explore the nature and scope of public recognition and/or lack of recognition (i.e., misrecognition) of STEM engagement experienced by undergraduate students at a Hispanic Serving Institution (HSI) prior to college. Our guiding research questions were,

1. In what ways do participants describe events where they experienced public recognition or misrecognition of their STEM-related engagement?
2. In what ways do participants associate these experiences with their STEM career aspirations or lack thereof?

**Research Design/Procedure**

This concurrent analysis of preliminary data is part of an ongoing phenomenological study that aims to understand the nature of STEM-related accomplishments, achievements, and barriers associated with underrepresented youth's STEM identities and career aspirations. Roulston (2010) stated that to conduct effective phenomenological interviews, the researcher or interviewer should be able to consider and ensure that the selected individuals have experience and can talk about their specific lived experiences. As such, the recruitment of interviewees was based on their responses to an NSF-funded survey study of racially and ethnically minoritized students' out-of-school-time STEM learning experiences. We purposefully recruited interviewees who indicated aspiring to either STEM or non-STEM careers in equal numbers, inviting them to an hour-long interview (van Teijlingen & Ireland, 2003). We began our interviews with a question about how they became interested in their current major (e.g., Dou & Cian, 2021) and followed up to inquire about experiences relevant to their career pursuits. As a part of the process of engaging with the data, the authors reviewed interviewees' survey responses prior to each interview, documenting their developing ideas using reflective memos both prior to and after each interview (Moore et al., 2021). After the first two interviews, we coded transcripts using descriptive codes aligned with our research questions and theoretical frameworks. The first and second authors met regularly to discuss the data and coding process with weekly input from the fourth author, who served as the project's principal investigator. We analyzed interview transcripts concurrently with data collection to test the durability of themes emerging from our discussions (e.g., Snowden & Martin, 2011).

The study participants were purposefully selected from a public university in Florida. The preliminary criteria for selecting participants were based on their aspirations to pursue STEM[1] and non-STEM[2] careers at the "end of high school" or "beginning of the first semester of college". Another criterion for the selection was based on the median score of their reported STEM identity score on the survey (see Table 1). We report on findings from our first five interviews; participant recruitment is ongoing at the time of the submission of this proposal, and we plan to interview 8-12 interviewees from each STEM and non-STEM participant group. These preliminary analyses are regarded as primarily contributing to further improvement of our interview protocol and testing the durability of our developing themes.

**Table 1**. Chart illustrating the sampling criteria and sample size

| | **Career Interest (1st level)** | **STEM Identity (2nd level)** | **Pseudonyms** |
|---|---|---|---|
| **Recruitment Criteria** | STEM Career Aspiration | High | Will, Sam, Penny |
| | | Low | - |
| | non-STEM Career Aspiration | High | Mercy |
| | | Low | Jessica |

## Analyses and Findings

Although Moustakas (1994) recommends that phenomenological research data should ideally be presented extensively, due to proposal requirements, we will focus on a few ideas organized into two categories of preliminary themes from our phenomenological analyses: (1)

[1] Science, Technology, Engineering and Mathematics (STEM) Careers include medical doctors, health professionals, life scientists, astronomers, earth/environmental scientists, physical scientists, computer scientists, engineers, mathematicians/statisticians, and science or math teachers.

[2] Non-STEM Careers include other teachers, anthropologists/archaeologists, social scientists, humanities professionals, visual/performing artists, business persons, lawyers, politicians, athletes/coaches, military personnel, and other non-STEM related careers.

the nature and scope of publicly recognized or misrecognized STEM engagement, and (2) how interviewees associated these experiences with their sense of recognition from others as a STEM person and their career aspirations.

**Description of STEM-Related Recognition and Misrecognition Events**

We explored how non-STEM participants describe the public recognition or misrecognition of their STEM engagement. Descriptions of these events shared a variety of similar characteristics. Many were related to recognition received in formal K-12 schooling contexts, such as public awards or acknowledgement of achievements, but some were related to public recognition received in informal learning contexts. For example, Will, an African American male and a freshman majoring in Computer Science, participated in out-of-school coding and web development classes since middle school and earned several certificates as accreditation and acknowledgement of his abilities from his teachers, specifically for developing his first website while in high school. When we asked him the reason why these were meaningful to him, he said,

> "I would say…anything that you may receive a certificate or award for or like programs [website] that you would make or like maybe a little game that you would code or something like that [Pause] or did anything that - that would be noteworthy by like teachers and [friends]."

For many of our interviewees, regardless of the context of the public recognition they received, their notion of the meaningfulness of that recognition was associated with strong emotions, particularly a sense of pride. Mercy, a Peruvian-American female pursuing a non-STEM major, described meaningful accomplishments as events that make her feel proud and good, such as receiving good grades in middle school. Relatedly, misrecognition events were also associated with strong emotions, but rather than a sense of pride, students described feelings of isolation and rejection. For example, while describing their experiences, both Sam, a non-binary Latine STEM major and Mercy, a non-STEM major, described receiving little or no recognition from their STEM teachers. For example, Mercy said she often struggled with her math performance in middle school, stating that her teacher never cared about her struggles, which she believed significantly hindered her from pursuing any subject that relies on math. She also felt less motivated to join STEM clubs due to a feeling of isolation (i.e., her friends preferred reading literature instead). Sam said that although she was engaged in math classes until high school, she felt less connected to them when one of her high school teachers seemed to favor a few students over her and others by overlooking her competence and excellence in the class. She described that this experience led her to feel less connected to mathematics and subjects requiring knowledge of mathematics.

**Association with STEM Identity and Career Choice**

Among our interviewees pursuing a STEM major, all reported that participating in STEM activities had significantly shaped their career aspirations. On the other hand, non-STEM career aspirants were more likely to describe events where they were generally recognized as good students (e.g., Principal’s Honor Roll). For example, one of our non-STEM participants, Jessica, a Cuban white female freshman pursuing a non-STEM major, when considering meaningful accomplishments that made her feel proud, said, “I always had straight A's. I would always get Principal’s Honor Roll. And so, this was normal for me, to participate in these things.” In this excerpt, Jessica speaks about a recognition of her high GPA across all subject areas. Jessica’s commentary did not speak specifically to STEM when asked about what she considers meaningful accomplishments.

Our non-STEM interviewees acknowledged being recognized for their STEM competence (e.g., getting awards, certificates, or invitations to participate in STEM competitions); however, their participation in STEM activities was primarily driven by external motivations (e.g., high school requirements) rather than intrinsic interests. As such, their narratives suggested that public recognition of STEM accomplishments, without a personal interest in STEM fields, may not meaningfully nurture an increase in STEM identity. For example, Jessica described having a strong support system, high math competency, several STEM-related accomplishments (e.g., awards and certificates), and never experienced any misrecognition, yet she never considered herself a STEM person.

The most significant misrecognition events students described experiencing were associated with formal mathematics education. Among all our interviewees, one of the non-STEM participants, Mercy, wanted to choose a career in STEM (i.e., veterinarian, Mathematician/Statistician, Science or Math teacher) from the beginning of middle school until the beginning of high school. Mercy reported that a lack of support and recognition from math teachers inspired her to choose a non-STEM career. This shift is a potential decay of her performance–competence in STEM that might have been fueled by the lack of external support and recognition. However, this lack of recognition in formal schooling contexts was, in some cases, mitigated by recognition and support from family.

## Contribution & Implications

Research has consistently delineated that STEM identity plays a key role in underrepresented youth's success in STEM fields (Carlone & Johnson, 2007; Dou & Cian, 2021), and "recognition" of STEM performances is a positive approach by teachers and parents to acknowledge youth's accomplishments (e.g., Mann, 1988), which can positively impact one's STEM identity. Our preliminary analysis suggests that undergraduate students, regardless of major, recall particular recognition and misrecognition events in STEM contexts that meaningfully shaped their STEM identity. These events occurred in formal and/or informal learning settings and were associated with strong emotions—a sense of pride or a sense of isolation, depending on the nature of the (mis)recognition. Moreover, they attributed these events as directly shaping their identification with STEM and their career choices, with recognition events having a positive effect and misrecognition events having a negative effect. However, recognition events in STEM contexts seemed to be only meaningful when the interviewee also described a personal interest or positive disposition toward STEM, and in some cases, misrecognition events were mitigated by family members.

The results of this study highlight the importance of further exploring the recognition and misrecognition events experienced by individuals seeking STEM and non-STEM degrees, especially those historically underrepresented in STEM fields, to better support their identification with STEM (e.g., Simpson & Bouhafa, 2020; Avraamidou, 2022). These insights and findings of the study may benefit NARST members who are concerned with the consequences of and solutions to racial and gender disparities in the STEM education field, particularly those whose research aims and involves designing effective interventions to develop a strong STEM identity among the populations who are traditionally underrepresented in STEM (e.g., women and ethnic minorities). The study's insights can also be applied to designing and developing formal and informal programs that aim to recognize youths' various STEM-related accomplishments while fostering personal interest in STEM fields. In addition, it may also benefit NARST members to conduct further research on reframing and reforming existing recognition approaches in various formal and informal STEM learning contexts to ensure a robust STEM identity development and STEM career aspiration among minority populations.